\documentclass[final]{elsart3p}

\usepackage{graphicx}

 \journal{Comp. Condens. Matter}
\usepackage{dcolumn}
\usepackage{bm}

\begin{document}

\begin{frontmatter}

\title{Electronic and gap properties of Sb and Bi based
halide  perovskites: An \textit{ab-initio} study}

\author{Athanasios Koliogiorgos, Sotirios Baskoutas and Iosif Galanakis}
\address{Department of Materials Science, School of Natural
Sciences, University of Patras,  GR-26504 Patra, Greece}

\begin{abstract}
Halide perovskites are currently under intense investigation due
to their potential applications in optoelectronics and solar
cells. Among them several crystallize in low symmetry lattice
structures like trigonal, hexagonal, orthorhombic and monoclinic.
Employing \textit{ab-initio} electronic structure calculations in
conjunction with generalized gradient approximation and hybrid
functionals we study a series of perovskites with the formula
A$_3$B$_2$X$_9$ which have been grown experimentally. A stands for
a monovalent cation like Cs, Rb, K or the organic methylammonium
molecule (MA), B is Sb or Bi, and X is a halogen. Moreover we
include in our study both the effect of spin-orbit coupling in the
halide perovskites and the influence of the orientation disorder
of the MA cation on the energy band gaps of these compounds. Most
compounds under study exhibit absorption in or close to the
optical regime and thus can find application in various
optoelectronic devices. Our results pave the way for further
investigation on the use of these materials in technology relevant
applications.
\end{abstract}


\end{frontmatter}

\maketitle

\section{Introduction}\label{Intro}

The continuous growth of energy consumption and the finite amount
of available fossil fuels has triggered the research on
alternative energy sources and especially the so-called "renewable
energy sources". Among them sun is the most promising since it is
an infinite energy sources and solar cells are under intense
study. The most widely used material for solar cells is silicon
with a history of over 60 years \cite{Green1993}. But the search
for new cheap, earth-abundant materials, which can substitute
silicon in solar cells, is at a peak in recent years
\cite{Yang2016}. Among the proposed materials, perovskites
\cite{Science1} consist a promising family of materials for the
photovoltaic (PV) industry \cite{Science2,Chen2015,Gratzel2014}.
Research on perovskites for PVs has still to address multiple
issues like efficiency (PCE), the width of the energy band gap,
that needs to correspond roughly to the visible spectrum, and
device stability in ambient conditions prior to their commercial
use. The interest on the family of perovskites also embraces other
technological important research regions like optoelectronic
devices \cite{Science3} and catalysis \cite{Science4}.

Early perovskites contained oxygen and had the general structure
of ABO$_3$. To achieve charge neutrality, A has to be a cation of
+2 valence and B a cation of +4 valence of dissimilar size like in
CaTiO$_3$ \cite{Pena2001}. Latter it was found that in the same
family of compounds an increasing number of materials can be
categorized \cite{Chen2015,Gratzel2014}. First, one can use
halogen atoms instead of oxygen, giving birth to the so-called
halide perovskites \cite{Hoefler2017} Charge neutrality implies
now that A is a monovalent cation and B a divalent.The A cation
does not have to be a pure element but also an organic molecule
and in this case the materials are known as hybrid or
organometallic halide perovskites
\cite{Yang2016,Gratzel2014,Hoefler2017,Papavassiliou2012}. The
choice of the two cations and the halogen atoms offers the ability
to tune the photoconductive properties \cite{Weber1978}. Among the
ABX$_3$ hybrid halide perovskites  is MAPbI$_3$, where MA stands
for the methylammonium cation CH$_3$NH$_3$$^+$, which has
attracted most of the attention due to its high charge-carrier
mobility and its absorption at optical regime
\cite{Yang2016,Hohnston2015,Brittman2015,Frost2014,Filippetti2014,Albero2016,Zhao2015,Zhou2016,Quarti2016}.

First-principles, also known as ab-initio, electronic structure
calculations are a powerful tool to study and predict the
properties of compounds and thus play a decisive role in modern
Materials Science. Several first-principles calculations have been
devoted to the study of MAPbI$_3$
\cite{Brivio2013,Walsh2015,Leguy2015,Mosconi2013,Motta2015,Koliogiorgos2017}.
The toxicity of the lead atoms led to the search for alternative
hybrid halide perovskites \cite{Stoumpos2014,Hao2014}, and
first-principles electronic structure calculations have been
devoted to the study of MABX$_3$ compounds where B a divalent
cation other than Pb and X an halogen atom
\cite{Ma2012,Bernal2014,Borriello2008,Jacobsson2015}. In a recent
publication, we studied extensively the electronic properties of
cubic MABX$_3$ compounds, where the divalent cation B was Ca, Sr,
Ba, Zn., Cd, Hg, Ge, Sr or Pb and the halogen atom X could be F,
Cl, Br or I  including also the case of mixed halide perovskites
\cite{Koliogiorgos2017}. Our results suggested that there were
some compounds like MAGeCl$_3$ and MAGeBr$_3$ which could replace
MAPbI$_3$ in photovoltaics in the future, but their experimental
growth in the cubic lattice structure is yet to be demonstrated.

Searching for the other possible replacements for the iodine
methylammonium lead perovskite, we extend our ab-initio study to
perovskites which do not crystallize in the cubic or pseudo-cubic
structure, but other low-symmetry lattices like trigonal,
hexagonal, orthorhombic or monoclinic  \cite{Hoefler2017}. In
these perovskites the stoichiometry changes and the general
chemical formula is now A$_3$B$_2$X$_9$, where A is a Cs, Rb, K or
methylammonium (MA) monovalent cation, B is a Bi or Sb trivalent
cation, and X is an I, Br or Cl monovalent anion. Due to the large
number of possible low-symmetry structures we have included in our
study compounds for which the lattice structure is known
experimentally. Our results show that the value of the energy band
gaps vary greatly among the studied materials and thus
A$_3$B$_2$X$_9$ perovskites can be promising resource of materials
for photovoltaic applications. In section 2 we present shortly the
details of our calculations, since they are similar to the ones in
reference \cite{Koliogiorgos2017} and the structural details of
the compounds under study. In section 3 we present and discuss our
results, while sections 4 and 5 are devoted to the effect of
spin-orbit coupling and the hybrid perovskites, respectively.
Finally in section 6 we summarize and present our conclusions.

\begin{table*}
  \caption{A$_3$B$_2$X$_9$ compounds with their respective crystallographic
  structure and the reference for the experimental data.
  We present also  the k-grid (as symmetry lowers it becomes less dense), the
energy cutoff ($E_{cut-off}$) in eV and the calculated energy band
gap values, $E_g$, using both the PBEsol and HSE06 functionals; in
parenthesis the $E_{cut-off}$ and $E_g$ values taking into account
also the spin-orbit coupling (SOC) effect }
  \label{table1}
  \begin{tabular}{lllllll}
    \hline

Compound  & Space group & \textbf{k}-points & $E_{cut-off}$ (eV) &
$E_g^\mathrm{PBEsol}$ (eV) & $E_g^\mathrm{HSE06}$ (eV\\
\hline

Cs$_3$Sb$_2$I$_9$ \cite{Peresh2011} &  Trigonal P3m1          & 5x5x2 & 300(140)  &  1.28(1.18) & 1.82\\

Cs$_3$Sb$_2$Br$_9$ \cite{Peresh2011} &  Trigonal P3m1         & 5x5x2 & 300(150)  &  1.40(1.31) & 2.07\\

Cs$_3$Sb$_2$Cl$_9$\cite{Timmermans2011} &  Trigonal P321      & 5x5x2 &  300(170)  &  2.12(2.07) & 2.97\\

Cs$_3$Bi$_2$I$_9$ \cite{Chabot1978}  & Hexagonal P63/mmc & 4x4x2 &
190(100)  &  1.89(0.40) &
2.56(0.88)\\

Cs$_3$Bi$_2$Br$_9$ \cite{Peresh2011} &  Trigonal P3m1         & 5x5x2 &  300(160)  &  2.46(1.63) & 3.08\\

Cs$_3$Bi$_2$Cl$_9$ \cite{Timmermans2011} &  Orthorhombic Pmcn & 2x5x4 & 150(100)  &  2.80(1.08) & -\\

Rb$_3$Sb$_2$I$_9$ \cite{Peresh2011} &  Monoclinic Pc          & 4x6x2 & 110(70)   &  1.66(1.57) & -\\

Rb$_3$Sb$_2$Br$_9$ \cite{Peresh2011} &  Trigonal P3m1         &
5x5x2 & 300(200)  &  1.55(1.45) & 2.21(1.98)\\

Rb$_3$Bi$_2$I$_9$  \cite{Sidey2000}&  Monoclinic Pc           &  4x6x2 & 100(70)   &  1.04(1.27) & -\\

Rb$_3$Bi$_2$Br$_9$ \cite{Peresh2011}  &  Orthorhombic Pnma    & 2x5x4 &150(90)   &  2.41(0.13) & -\\

K$_3$Sb$_2$I9 \cite{Preitschaft2004}      & Monoclinic P21/n   &
4x5x2 & 130(80)   &  0.68(0.48) & - \\

K$_3$Bi$_2$I9 \cite{Lehner2013}     & Monoclinic P21/n        &  4x5x2 & 120(80)   &  0.68(0.64) & - \\

MA$_3$Sb$_2$I$_9$  \cite{Eckhardt2016}&  Hexagonal P63/mmc    & 5x5x2 & 180(100)  &  1.73(1.52) & 2.36\\

MA$_3$Bi$_2$I$_9$  \cite{Eckhardt2016}&  Hexagonal P63/mmc    & 5x5x2 & 180(100)  &  2.19(0.45) & 2.85\\
   \hline

  \end{tabular}
\end{table*}

\section{Computational method}\label{CompDet}

To perform the calculations we employed the Vienna Ab-initio
Simulation Package (VASP) \cite{VASP}. VASP is a powerful program
for performing ab-initio quantum mechanical molecular dynamics
simulations, developed by the Institut fur Metaliphysik of the
Universitat Wien \cite{VASP}.In this study, similarly to the one
in reference \cite{Koliogiorgos2017}, we employed the projector
augmented planes (PAW) pseudopotentials in reciprocal space to
perform the simulations \cite{PAW}. VASP also allows for
self-consistent  calculations including the spin-orbit coupling
(SOC).

As charge density functionals, we employed, first, the Generalized
Gradient Approximation (GGA) as parameterized by Perdew, Burke and
Ernzerhof for solids known as the PBEsol approximation
\cite{PBE,PBEsol}. We used the PBEsol results to perform also
calculations with the Hunde, Scuseria and Ernzerhof hybrid
functional known as HSE06 \cite{HSE06,VASP-HSE06}. The latter
accounts for the exchange energy in a semi-empirical way
considering it to be a mixture of the usual GGA exchange and of
the exact Hartree-Fock exchange energy. The formula used is :
\begin{equation}
E_{xc} = \frac{1}{4} E_x^{exact} + \frac{3}{4} E_x^{GGA} +
E_c^{GGA}
\end{equation}
where the $E_x^{exact)}$ is the Hartree-Fock contribution and
$E_x^{GGA}$ and $E_c^{GGA}$ are the respective exchange and
correlation energy contributions by the GGA approximation, i.e.
the PBEsol functional. Due to its nature HSE06 tends to restore
the correct values of the energy band gaps which are
underestimated by usual GGA calculations as shown also in the case
of cubic perovskites \cite{Koliogiorgos2017}. Thus, the HSE06 is
an accurate functional that can be very useful where the
calculation of the optical band gap is the main concern. A
drawback of this method is that it requires cpu time orders of
magnitude larger that the usual GGA functionals. As a result we
were able with our available computer resources to achieve
convergence using HSE06 only for the compounds with the trigonal
and hexagonal structures; calculations for the orthorhombic and
monoclinic structures did not converge. Convergence was even more
difficult when we included also the SOC and the combined HSE06 and
SOC calculations converged only for the trigonal
Rb$_3$Sb$_2$Br$_9$ and hexagonal Cs$_3$Bi$_2$I$_9$ compounds.

\begin{figure*}
\includegraphics[width=\textwidth]{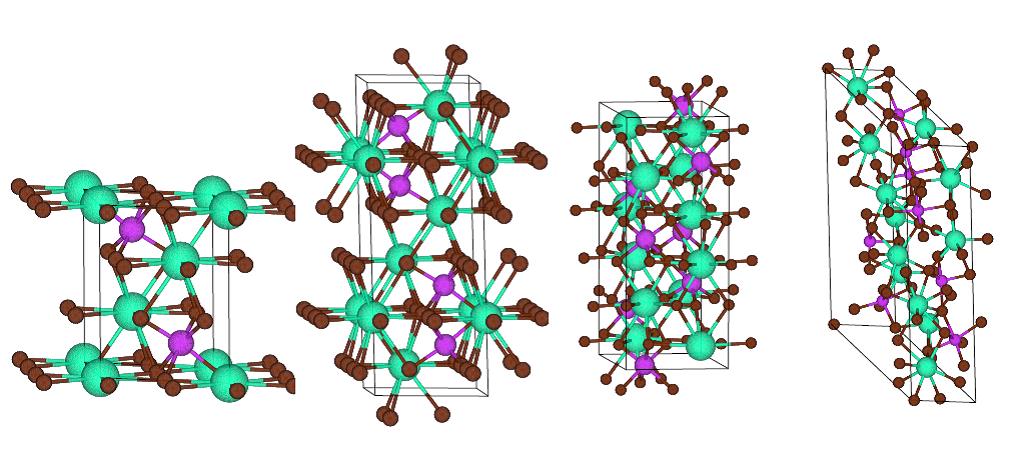}
\caption{Structure of A$_3$B$_2$X9 perovskites (going from left to
right): (a) trigonal structure, space group P3m1; (b) hexagonal
structure, space group P63/mmc; (c) orthorhombic structure, space
group Pmcn; (d) monoclinic structure, space group Pc. A atoms are
in green, B atoms are in purple and X atoms in brown. }
\label{fig1}
\end{figure*}

As we will also discuss later when we will present our results an
important parameter in the electronic band structure calculations
is the cut-off energy which is the maximum kinetic energy of the
plane waves used in the calculations. The larger is this value,
the most accurate are the calculations. As the symmetry of the
lattice lowers going from the trigonal to the monoclinic lattice,
the energy cut-off used for the calculations has to lower in order
to achieve convergence. Moreover the inclusion of SOC leads to an
extra lowering of the possible cut-off energy. In table
\ref{table1} we present both the cut-off energy, $E_{cut-off}$ in
eV, as well as the \textbf{k}-points grid in the reciprocal space
used in the calculations; we have used for the latter the
Monkhorst-Pack scheme \cite{Monkhorst}. Test calculations with
denser grids yielded roughly the same results establishing the
accuracy of our results.

In table \ref{table1} we present all the compounds which we have
studied together with the lattice structure and the reference to
the experimental structural details. The monovalent cation is Cs,
Rb or K and the trivalent cation is Sb or Bi, while for the
halogen Cl, Br or I are present. We have restricted ourselves to
compounds for which experimental evidence exists regarding their
lattice. We have also included in our study the case of the hybrid
MA$_3$Bi$_2$I$_9$ and MA$_3$Sb$_2$I$_9$ compounds crystallizing in
the hexagonal structure similar to Cs$_3$Bi$_2$I$_9$ as shown in
reference \cite{Eckhardt2016} for MA$_3$Bi$_2$I$_9$; for the
Sb-based compound we assumed that it crystallizes in the same
lattice as the Bi-based one. As can be seen in the table, the
structures of the compounds fall into four space group families
(going from higher to lower symmetry): trigonal, hexagonal,
orthorhombic and monoclinic.  Examples of the A$_3$B$_2$X$_9$
basic lattice structures are shown in figure \ref{fig1}(a-d). As
can be seen from the pictures, as we progress from higher to lower
symmetries, the unit cell becomes larger and thus the increased
number of atoms in the unit cell leads to a large increase in the
required computer power and in the computational time.

\begin{figure}
\includegraphics[width=\columnwidth]{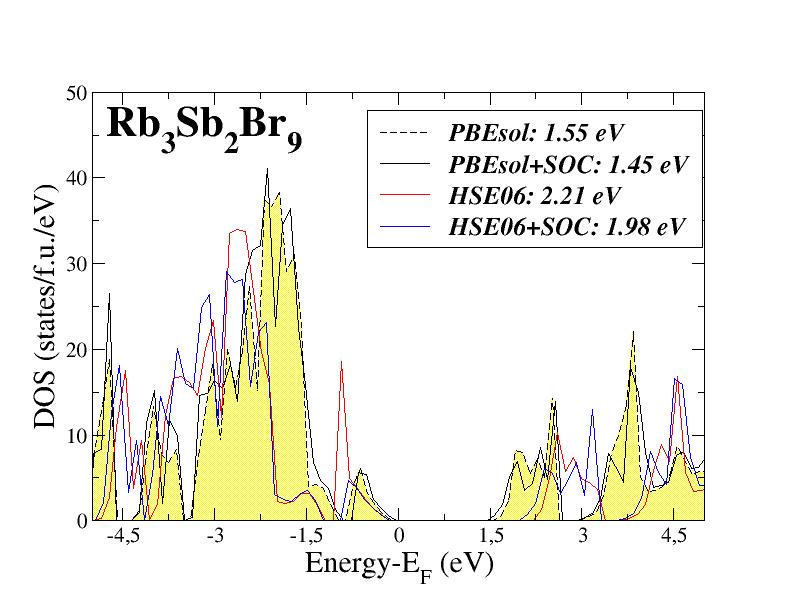}
\includegraphics[width=\columnwidth]{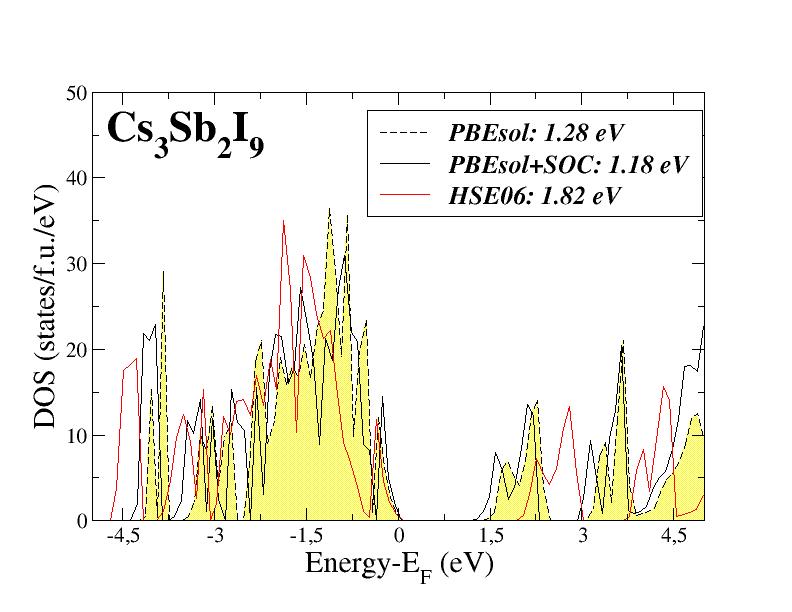}
\caption{Density of states for two A$_3$B$_2$X$_9$ compounds,
Rb$_3$Sb$_2$Br$_9$ (upper) and Cs$_3$Sb$_2$I$_9$ (lower panel)
using PBEsol and HSE06 functionals, and with or without SOC
effect. The valence and conduction bands, as well as the band gap
between them, are visible. The k-point grid and the energy cutoff
are presented  in table \ref{table1}. The zero energy in the
horizontal axis is set to be the Fermi level.} \label{fig2}
\end{figure}

\section{Halide perovskites}\label{Halide}

As discussed above our aim is to study the   perovskite compounds
with respect to their electronic properties, and especially their
band gap, in order to discern which of them would be suitable for
optoelectronic applications, such as solar cells. In table
\ref{table1}, we present the results produced using the PBEsol and
HSE06 functionals first without taking into account SOC; the
effect of the SOC will be discussed in the next section.

The first family of A$_3$B$_2$X$_9$ which we studied are the ones
where A is Cs. For all compounds under study we have used the
experimental lattice structure from the references in table
\ref{table1}. The compounds where the trivalent cation is Sb
crystallize all in the trigonal lattice. For this lattice
structure we used a cut-off energy of 300 eV and thus band gaps
are expected to be highly accurate. Within PBEsol the width of the
gap ranges between 1.28 eV for the iodide compound up to 2.12 eV
for the chloride compound. When HSE06 was employed instead of
PBEsol the values of the obtained gaps increase and reach 1.82,
2.07 and 2.97 eV for the compounds containing as halogen atom I,
Br and Cl, respectively. This behavior is similar to the one in
cubic halide compounds studied in reference
\cite{Koliogiorgos2017} and the oxygen perovskites
\cite{Franchini2014}. The tendency observed with the halogen atom
present in the compound can be easily explained by the density of
states (DOS) presented in figure \ref{fig2} for Cs$_3$Sb$_2$I$_9$.
The valence band is made up of the halogen $p$-states and the
conduction band from the Sb $p$-states. Thus the gap opens due to
the $p-p$ hybridization mechanism.This ensures that the gap values
are close or within the optical regime contrary to cases where the
gap opens due to a $p-d$ type hybridization and the gap is much
larger \cite{Koliogiorgos2017}. Secondly as we move from Cl to Br
and then to I, the valence $p$-states are located higher in energy
narrowing the gap. HSE06 with respect to PBEsol shifts the
conduction band slightly higher in energy opening the gap without
influencing the shape of the bands.

In the Cs-based compounds when instead of Sb we have the heavier
isovalent Bi, the three resulting compounds have no more the same
lattice structure. Cs$_3$Bi$_2$I$_9$ is hexagonal, the
Cs$_3$Bi$_2$Br$_9$ is trigonal and Cs$_3$Bi$_2$Cl$_9$ ir
orthorhombic. As mentioned above for the hexagonal and
orthorhombic structures we had to use a considerably smaller
cut-off energy with respect to the trigonal one, which narrows the
accuracy and reliability of the results obtained for these lower
symmetry systems. Moreover in the case of the orthorhombic
Cs$_3$Bi$_2$Cl$_9$ compound we were not able to converge the HSE06
calculations. PBEsol yielded band gaps of 1.89, 2.46 and 2.80 eV
for the I-, Br- and Cl-based compounds respectively. HSE06 further
increased the values of the band gap to 2.56 and 3.08 eV for the I
and Br compounds, respectively. These values exceed the optical
regime. The origin of the gap is similar to the Sb-based compounds
discussed just above.

The next step is our study is to replace Rb for Cs.
Experimentally, only the compounds with Br and I as halogen have
been grown and the lattices vary. Only Rb$_3$Sb$_3$Br$_9$ is
trigonal with a PBEsol band gap of 1.55 eV and a HSE06 band gap of
2.21 eV. For the other three compounds with the lower symmetry we
were not able to converge the HSE06 calculations. Finally, we have
to mention that although PBEsol gives for Rb$_3$Bi$_3$I$_9$ a band
gap of 1.04 eV and thus one could expect HSE06 to give a value
within the optical regime, the relatively small value of the
energy cut-off of 100 eV due to its monoclinic lattice structure
makes the calculated values less trustworthy than for the trigonal
compounds and deviations from experiments can occur. Finally, we
have also studied the K$_3$Sb$_2$I$_9$ and K$_3$Bi$_2$I9
compounds. Both crystallize in the monoclinic lattice and PBEsol
yields band gaps of about 0.7 eV much smaller that the optical
spectra.

For reasons of completeness we have included in table \ref{table1}
also the case of the hybrid MA$_3$Sb$_2$I$_9$ and
MA$_3$Bi$_2$I$_9$ compounds; the latter exists experimentally
while for the former we have used the same hexagonal lattice
structure. The cut-off energy for our calculations is 180 eV and
we have sued a 5$\times$5$\times$2 \textbf{k}-points grid in the
reciprocal space. The obtained HSE06 gaps are exceeding the 2 eV
reaching a value of 2.85 eV for the MA$_3$Bi$_2$I$_9$ compound. In
section \ref{Hybrid} we will discuss in detail the influence of
the disorder regarding the orientation of the MA cations on the
obtained band gaps.

\begin{figure}
\includegraphics[width=\columnwidth]{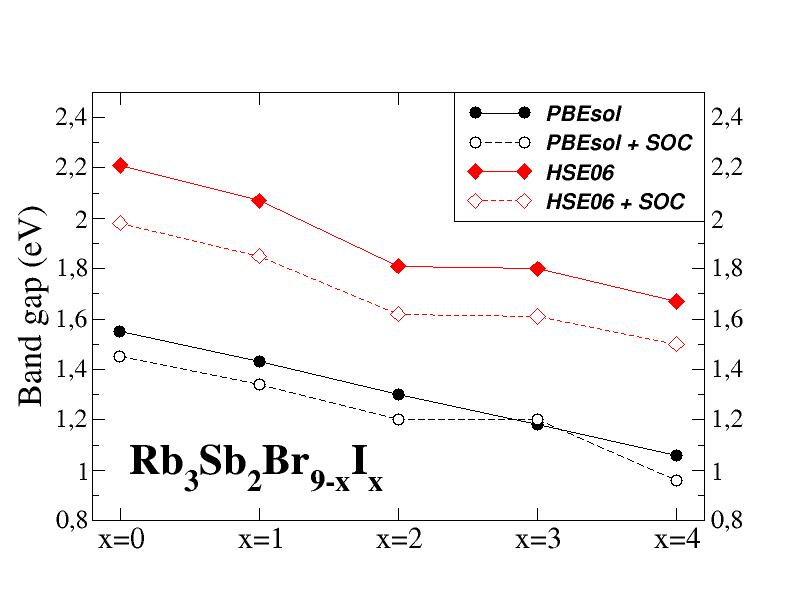}
\caption{Band gaps in eV of the mixed Rb$_3$Sb$_2$Br$_{9-x}$I$_x$
perovskites, with x$\leq$4. The space group for the mixed
compounds is assumed to be the same as for Rb$_3$Sb$_2$Br$_9$,
i.e. trigonal P3m1. The energy cutoff is 300 eV without SOC and
200 eV with SOC. The k-point grid is 5x5x2. For the HSE06+SOC case
we present the expected values via extrapolation (see text for
details). } \label{fig3}
\end{figure}

\subsection{Effect of spin-orbit coupling}\label{SOC}

We proceeded to calculate the electronic properties of the
compounds taking into account SOC using the VASP program. Previous
studies suggested that spin-orbit coupling has a strong effect on
the lead halide perovskites, reducing the band gap by 0.3 up to
roughly 1 eV \cite{Pedesseau2014}, while the effect on lead-free
perovskites was found to be weaker \cite{Bernal2014}. Thus it
seems that the influence of SOC is materials specific. In table
\ref{table1} we have included in parenthesis the energy band gap
including SOC as well as the energy cut-off used in the
calculations. We we able to converge the SOC calculations using
PBEsol for all compounds but we achieved convergence using HSE06
only in the case of Cs$_2$Bi$_2$I$_9$ and Rb$_3$Sb$_2$Br$_9$
compounds. A first glimpse at the results suggests that SOC is
very important and there are cases like Cs$_2$Bi$_2$I$_9$ where
the inclusion of SOC decreases the PBEsol band gap value by 79\%
and the HSE06 band gap by 66\% . But these value are not really
trustworthy since the inclusion of SOC implies a large decrease of
the energy cut-off used in the calculation decreasing the accuracy
of the calculations.

The prototype case to study the effect of SOC is
Rb$_3$Sb$_2$Br$_9$ since the energy cut-off in that case is 300 eV
without SOC and 200 eV with SOC, which should lead to accurate
results in both cases. The band gap in this cases decreases by 0.1
eV from 1.55 to 1.45 eV in the case of PBEsol when the SOC is
included, and from 2.21 to 1.98 eV in the case of the HSE06
calculations when the SOC is included. Thus as expected SOC has a
non-negligible effect on the band gap lowering it by about 10.5\%
in the case of the HSE06 functional. In the upper panel of figure
\ref{fig2} we present the DOS for all four cases under study to
investigate the effect of SOC. SOC slightly moves the bands both
in the case of PBEsol and HSE06 calculations without considerably
affecting the shape of the initial bands without SOC.

In order to see if the decrease of the band gap is sensitive on
the halogen atoms, we have partially substituted I for Br in
Rb$_3$Sb$_2$Br$_9$ resulting in the Rb$_3$Sb$_2$Br$_{9-x}$I$_x$
keeping the same trigonal structure. We have taken into account up
to $x=4$ since for larger values we are closer to
Rb$_3$Sb$_2$I$_9$ which crystallized in the monoclinic lattice. We
have performed calculations with and without SOC using the PBEsol
functional and present our results in figure \ref{fig3}. The
calculated band gap values with SOC are about 0.1 eV smaller than
the values without SOC with the exception of $x=3$ where values
are similar. Thus the effect of spin-orbit coupling does not
really depend on the halogen atom but on the trivalent cation and
the lattice structure. We have also made calculations using HSE06
without SOC and we present these results in figure 3 together with
predicted values with SOC assuming that the SOC reduces the band
gap by 10.5\%. The substitution of I for Br tunes the band gap
which now falls within the optical regime and thus the
Rb$_3$Sb$_2$Br$_{9-x}$I$_x$ compounds are suitable for solar cell
applications.

\begin{figure}
\includegraphics[width=\columnwidth]{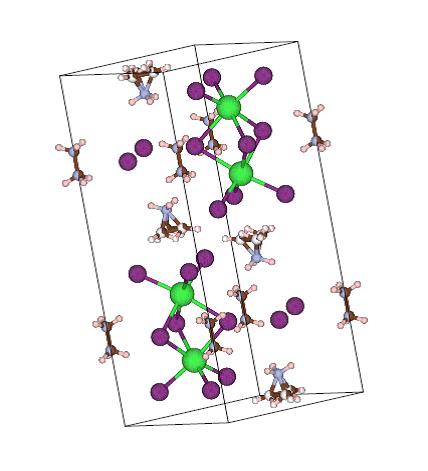}
\caption{MA$_3$Bi$_2$I$_9$ unit cell in hexagonal structure. The
atoms in the unit cell are: Bi in green, I in purple, C in brown,
N in blue and H in pink. CH$_3$NH$_3$$^+$ cations are positioned
on the edges of the unit cell and inside the unit cell, with
varying orientations. } \label{fig4}
\end{figure}

\subsection{The case of hybrid halide perovskites}\label{Hybrid}

As mentioned before, the A cation in the A$_3$B$_2$X$_9$ formula
can be inorganic or organic. In the case of organic A, the
compound is a hybrid halide perovskite. A case in point that has
been studied in literature is MA$_3$Bi$_2$I$_9$ perovskite where
MA stands for CH$_3$NH$_3$$^+$) known widely as the methylammonium
cation \cite{Eckhardt2016,Hoye2016}.

The crystallographic structure of MA$_3$Bi$_2$I$_9$ has been found
to be the hexagonal structure with two formula units per unit
cell. Using the experimental data, we constructed the perovskite
unit cell that was used in the simulation program, as seen in
figure \ref{fig4}. As can be seen in the figure, the compound
shows disorder with respect to the orientation of MA. We can
distinguish two types of disorder. The two CH$_3$NH$_3$ cations on
the side of the unit cell show a disorder with respect to the
position of the carbon and nitrogen atoms which can exchange
positions. In the case of the four inner CH$_3$NH$_3$ cations, the
nitrogen atoms are fixed and the carbon atoms can have three
different positions forming a fictional pyramid. Thus the question
rises whether the relative positions of the CH$_3$NH$_3$ cations
in the unit cell affects the width of the gap.

\begin{table}
  \caption{PBEsol band gaps (in eV), in 10x10x4 k-point grid, for MA$_3$Bi$_2$I$_9$
  for different MA arrangements (see text for detals). }
  \label{table2}
  \begin{tabular}{lllll}
    \hline
Inner &  \multicolumn{2}{c}{Outer arrangements} \\

arrangement & Parallel & Antiparallel \\ \hline

1st & 1.90  &  1.75 \\

2nd & 2.18  &  2.05 \\

3rd & 1.75  &  1.75 \\

4th & 2.04  &  2.17 \\

5th & 1.77  &  1.64 \\

6th & 1.89  &  1.90 \\ \hline

  \end{tabular}
\end{table}

In order to answer the question mentioned just above we carried
out electronic structure calculations using the PBEsol and
assuming 12 different variations of the distribution of the
orientation of the MA cations in the unit cell. Our results are
summarized in table \ref{table2}. For the MA cations in the inner
part of the unit cell we considered 6 different possible
arrangements. For the MA cations along the sides of the unit cell
we considered two cases: assuming the consecutive MA cations to
have either parallel arrangement (for both the C atoms is located
below the N atoms) or antiparallel arrangement. We found out that
for the hybrid perovskites a larger \textbf{k}-points grid is
needed with respect to the usual halide perovskites and thus for
the calculations we have used a 10$\times$10$\times$4 grid. This
also explains the difference with the results for the same
material in table \ref{table1} which corresponds to the first case
presented in table \ref{table2}. As can be seen the band gaps are
actually varying between the 12 studied cases ranging from 1.64 eV
to 2.18 eV. This is a large variation of more than half eV and may
make the material suitable or not suitable for applications where
absorption in a specific range of frequencies is needed. For the
same inner arrangement the difference between the values for the
parallel and antiparallel orientation of the outer MA cations is
smallest, being at most 0.15 eV for the 1st inner arrangement.
Thus the orientation disorder of the inner MA cations should
affect the obtained energy band gap in samples. We should finally
note that the DOS graphs reveal that the gap is formed between the
occupied $p$-states of the iodine and the empty $p$-states of Bi
similar to the usual halide perovskites studied above.

\section{Summary and conclusions}\label{Concl}

We studied using the program VASP for first-principle calculations
the structural, electronic and optical properties of 14 halide and
hybrid halide perovskites with the general formula A$_3$B$_2$X$_9$
and with trigonal, hexagonal, orthorhombic or monoclinic
structure, where A is a monovalent cation, B a trivalent cation
like Sb or Bi of dissimilar size and X a halogen. As the symmetry
of the lattice lowered, the calculated band gaps were determined
with smaller accuracy due to the lower energy cut-off used for the
kinetic energy of the plane waves. We employed the generalized
gradient approximation for all calculations as well as a more
accurate hybrid functional. Where possibly, we have included also
the spin-orbit coupling (SOC) in our calculations. We also studied
separately two hybrid halide compounds of the group, namely
MA$_3$Bi$_2$I$_9$, which exists experimentally, and
MA$_3$Sb$_2$I$_9$ assuming the same lattice structure, were MA
(CH$_3$NH$_3$$^+$) is the methylammonium organic cation.

All materials under study produced band gaps ranging between 1 and
3 eV, due to the $p-p$ hybridization responsible for the
appearance of the gap, and thus either are at the optical regime
or near it and can find applications in optoelectronic and solar
cell applications. The hybrid functional produced larger band gaps
as expected for the trigonal and hexagonal materials but we could
not get convergence for the lower-symmetry orthorhombic and
monoclinic lattice structures. We have also carried out
calculations including SOC but the low values for the energy
cut-off do not allow an accurate determination of its exact effect
although it leads to a significant shrinkage of the band gap. A
prototype case is Rb$_3$Sb$_2$Br$_9$ for which we carried out
calculations combining both the hybrid functional and SOC and for
which the latter led to a 10.5\%\ decrease of the band gap. For
this material we also substituted partially I for Br and the
resulting compounds have shown a decrease in the gap with the
increase of the I concentration but the effect of SOC was almost
independent of the I concentration.

In the case of hybrid halide compound MA$_3$Bi$_2$I$_9$, we
performed calculations assuming different orientations of the MA
cation. Our results suggest that the orientation of the
methylammonium cations in the unit cell plays a significant role
in the final band structure and band gap, and  a great variation
of the band gap was observed. As a result, we ascertain that, if
in the synthesis process an engineering of the orientation of the
organic molecules can be achieved, then a substantial tuning of
the band gap is possible  making the hybrid halide compounds a
promising candidate for solar cell technology.

Consequently, we can conclude that it is possible to search for
perovskites suitable for solar cell and optoelectronic
applications among the low-symmetry perovskites. But the width of
the energy band gap is materials specific and in the case of the
hybrid perovskites, the orientation of the organic molecules plays
a decisive role. We expect our results to intrigue further
ab-initio calculations as well as experiments on these promising
materials.

\ack{Authors acknowledge financial support from the  project
PERMASOL (FFG project number: 848929).}


\begin{thebibliography}{99}

\bibitem{Green1993}
M.A. Green,  Silicon solar cells: evolution, high-efficiency
design and efficiency enhancements, Semiconduct. Sci. Technol. 8
(1993) 1-12.


\bibitem{Yang2016}
L. Yang, A.T. Barrows, D.G. Lidzey, T. Wang, Recent processes and
challenges of organometal halide perovskite solar cells, Rep.
Prog. Phys. 79 (2016) 026501.

\bibitem{Science1}
K. Hirose, R. Sinmyo, J. Hernlund, Perovskite in Earth's deep
interior, Science 358 (2017) 734–738.

\bibitem{Science2}
J.-P. Correa-Baena, M. Saliba, T. Buonassisi, M. Gr\"atzel, A.
Abate, W. Tress, A. Hagfeldt, Promises and challenges of
perovskite solar cells, Science 358 (2017) 739–744.

\bibitem{Chen2015}
Q. Chen,  N. De Marco, Y. Yang, T. Song, C. Chen, H. Zhao, Z.
Hong, H. Zhou, Y. Yang,  Under the spotlight: The
organic-inorganic hybrid halide perovskite for optoelectronic
applications,  Nano Today 10 (2015) 355-396.


\bibitem{Gratzel2014}
M. Gr\"atzel, The light and shade of perovskite solar cells, Nat.
Mater.13 2014) 838-842.

\bibitem{Science3}
M.V. Kovalenko, L. Protesescu, M.I. Bodnarchuk, Properties and
potential optoelectronic applications of lead halide perovskite
nanocrystals, Science 358 (2017)745–750.

\bibitem{Science4}
J. Hwang, R.R. Rao, L. Giordano, Y. Katayama, Y. Yu, Y.
Shao-Horn1, Perovskites in catalysis and electrocatalysis, Science
358 (2017) 751–756.

\bibitem{Pena2001}
M.A. Pena, J.L.G. Fierro,Chemical structure and performance of
perovskite oxides, Chem. Rev. 101 (2001) 1981-2017.

\bibitem{Hoefler2017}
S.F. Hoefler, G. Trimmel, T. Rath, Progress on lead-free metal
halide perovskites for photovoltaic applications: a review,
Monatsh.  Chem.  148 (2017) 795.

\bibitem{Papavassiliou2012}
G.C. Papavassiliou, G. Pagona, N. Karousis, G.A. Mousdis, I.
Koutselas, A.  Vassilakopoulou, Nanocrystalline/microcrystalline
materials based on lead-halide units, J. Mat. Chem. 22 (2012)
8271-8280.

\bibitem{Weber1978}
D. Weber, Inst. Anorg. Chem. Univ. Stutt. 33B (1978) 1443-1445.

\bibitem{Hohnston2015}
M.B. Johnston, L.M. Herz, Hybrid Perovskites for Photovoltaics:
Change-Carrier Recombination, Diffusion, and Radiative
Efficiencies, Acc. Chem. Res. (2015). DOI:
10.1021/acs.accounts.5b00411.

\bibitem{Brittman2015}
S. Brittman, G.W.P. Adhyaksa, E.C. Garnett, The expanding world of
hybrid perovskites: materials properties and emerging
applications, MRS Commun. 5 (2015) 7-26.

\bibitem{Frost2014}
J. M. Frost, K.T.  Butler, F. Brivio, C.H. Hendon, M. van
Schilfgaarde, A.  Walsh, Atomistic origins of high-performance in
hybrid halide perovskite solar cells, Nano Lett. 14 (2014)
2584-2590.

\bibitem{Filippetti2014}
A. Filippetti, A. Mattoni,  Hybrid perovskites for photovoltaics:
Insights from first principles, Phys. Rev. B 89 (2014) 125203.

\bibitem{Albero2016}
J. Albero, A.M. Asiri, H. Garcia,  Influence of the Composition of
Hybrid Perovskites on their Performance in Solar Cells, J. Mater.
Chem. A (2016). DOI: 10.1039/C6TA00334F.

\bibitem{Zhao2015}
Y. Zhao, K. Zhu, K. Organic-inorganic hybrid lead halide
perovskites for optoelectronic and electronic applications, Chem.
Soc. Rev. (2015). DOI: 10.1039/c4cs00458b.

\bibitem{Zhou2016} Y.
Zhou, M. Yang, O.S. Game, W. Wu, J. Kwun,  M.A. Strauss, Y. Yan,
J. Huang, K. Zhu, N.P. Padture, Manipulating Crystallization of
Organolead Mixed-Halide Thin Films in Antisolvent Baths for
Wide-Bandgap Perovskite Solar Cells, ACS Appl. Mater. Interfaces 8
(2016) 2232–2237.


\bibitem{Quarti2016}
C. Quarti, E. Mosconi, J.M. Ball, V. D'Innocenzo, C. Tao, S.
Pathak, H.J. Snaith, A. Petrozza, F. De Angelis, Energy Environ.
Sci. 9 (2016) 155.




\bibitem{Brivio2013}
F. Brivio, A.B. Walker, A. Walsh, A. Structural and Electronic
 Properties of Hybrid Perovskites for High-Efficiency Thin-Film Photovoltaics
 from First-Principles, APL Mater. 1 (2013) 042111.

\bibitem{Walsh2015}
A. Walsh,  Principles of chemical bonding and band gap engineering
in hybrid organic-inorganic halide perovskites, J. Phys. Chem. C
(2015). DOI: 10.1021/jp512420b.

\bibitem{Leguy2015}
A.M.A. Leguy, J.M. Frost, A.P. McMahon, V.G.  Sakai, W.
Kockelmann, C. Law, X. Li, F. Foglia, A. Walsh, B.C. O'Regan, J.
Nelson, J.T.  Cabral, P.R.F.  Barnes, The dynamics of
methylammonium ions in hybrid organic-inorganic perovskite solar
cells, Nature Commun. 6 (2015) 7124.

\bibitem{Mosconi2013}
E. Mosconi, A. Amat, M.L.  Nazeeruddin, M.  Gr\"atzel, F.; De
Angelis,   First-principles modeling of mixed halide organometal
perovskites for photovoltaic applications, J. Phys. Chem. C 117
(2013) 13902-13913.

\bibitem{Motta2015}
C. Motta, F. El-Mellouhi, S.  Sanvito,  Charge carrier mobility in
hybrid halide perovskites, Nat. Sci. Reports 5 (2015) 12746.

\bibitem{Koliogiorgos2017}
A. Koliogiorgos, S. Baskoutas, I. Galanakis, Electronic and gap
properties of lead-free perfect and mixed hybrid halide
perovskites: An ab-initio study, Comput. Mater. Sci. 138 (2017)
92-98.

\bibitem{Stoumpos2014}
F. Hao, C.C.  Stoumpos, D.H. Cao, R.P.H. Chang, M.G. Kanatzidis,
Lead-free solid-state organic-inorganic halide perovskite solar
cells, Nature Photonics 8 (2014) 489-494.


\bibitem{Hao2014}
F. Hao, C. Stoumpos, D. Cao, R. Chang, M. Kanatzidis, Lead-free
solid-state organic-inorganic halide perovskite solar cells, Nat.
Phot. 8 (2014) 489-494.

\bibitem{Ma2012}
C. Ma, M. Brik, Hybrid density-functional calculations of
structural, elastic and electronic properties for a series of
cubic perovskites CsMF3 (M = Ca, Cd, Hg and Pb), Comput. Mater.
Sci. 58 (2012) 101-112.

\bibitem{Bernal2014}
C. Bernal, K. Yang, First-principles hybrid functional study of
the organic-inorganic perovskites CH$_3$NH$_3$SnBr$_3$ and
CH$_3$NH$_3$SnI$_3$, J. Phys. Chem. C 118 (2014) 24383–24388.

\bibitem{Borriello2008}
I. Borriello, G. Cantele, D.  Ninno, Ab initio investigation of
hybrid organic-inorganic perovskites based on tin halides, Phys.
Rev. B 77 (2008) 235214.

\bibitem{Jacobsson2015}
T.J. Jacobsson, M. Pazoki, A. Hagfeldt, T. Edvinsson,
Goldschmidt's rules and strontium replacement in lead halogen
perovskite solar cells: Theory and preliminary experiments on
CH$_3$NH$_3$SrI$_3$, J. Phys. Chem. C 119 (2015) 25673-25683.

\bibitem{VASP}
G. Kresse, J. Furthm\"uller,  Efficient iterative schemes for
\textit{ab initio} total-energy calculations using a plane-wave
basis set, Phys. Rev. B 54 (1996) 11169.

\bibitem{PAW}
J. Kresse, D. Joubert, From ultrasoft pseudopotentials to the
projector augmented-wave method, Phys. Rev. B 59 (1999) 1758.

\bibitem{PBE}
J.P. Perdew, K. Burke, M. Ernzerhof,  Generalized Gradient
Approximation Made Simple, Phys. Rev. Lett. 77 (1998) 3865.

\bibitem{PBEsol}
J.P. Perdew, A.   Ruzsinszky, G.I.  Csonka, O.A.  Vydrov, G.E.
Scuseria, L.A. Constantin, X.  Zhou, K. Burke, K. Restoring the
Density-Gradient Expansion for Exchange in Solids and Surfaces,
Phys. Rev. Lett. 100 (2008) 136406; \textit{ibid.} 102 (2009)
039902(E).

\bibitem{HSE06}
J. Heyd, G.E. Scuseria, M.   Ernzerhof, M. Hybrid functionals
based on a screened Coulomb potential, J. Chem. Phys. 118 (2003)
8207; \textit{ibid.} 124 (2006) 219906(E).

\bibitem{VASP-HSE06}
J. Paier, M. Marsman, K. Hummer, G.  Kresse, I.C.  Gerber,  J.G.
Angy\'an, Screened hybrid density functionals applied to solids,
J. Chem. Phys. 124 (2006) 154709; \textit{ibid.} 125 (2006)
249901E.

\bibitem{Monkhorst}
H.J. Monkhorst, J.D.  Pack, Special points for Brillouin-zone
integrations, Phys. Rev. B 13 (1976) 135188.

\bibitem{Eckhardt2016}
K. Eckhardt, V. Bon, J. Getzschmann, J. Grothe, F. Wisser, S.
Kaskel, Crystallographic insights into
(CH$_3$NH$_3$)$_3$(Bi$_2$I$_9$): a new lead-free hybrid
organic-inorganic material as a potential absorber for
photovoltaics, Chem. Commun. 52 (2016) 3058-3060.

\bibitem{Franchini2014}
C. Franchini,  Hybrid functionals applied to perovskites, J.
Phys.: Condens. Matter, 26 (2016) 253202.

\bibitem{Pedesseau2014}
L. Pedesseau J. Jancu, A. Rolland, E. Deleporte, C. Katan, J.
Even, Electronic properties of 2D and 3D hybrid organic/inorganic
perovskites for optoelectronic and photovoltaic applications, Opt.
Quant. Electron 46 (2014) 1225-1232.

\bibitem{Hoye2016}
R. Hoye, R. Brandt, A. Osherov, V. Stevanovic, S. Stranks, M.
Wilson, H. Kim, A. Akey, J. Perkins, R. Kurchin, J. Poindexter, E.
Wang, M. Bawendi, V. Bulovic, T. Buonassisi, Methylammonium
Bismuth Iodide as a Lead-Free, Stable Hybrid Organic-Inorganic
Solar Absorber, Chem. Eur. J. 22 (2016) 2605-2610.


\bibitem{Peresh2011}
E. Peresh, V. Sidei, O. Zubaka, I. Stercho,
K$_2$(Rb$_2$,Cs$_2$,Tl$_2$)TeBr$_6$(I$_6$) and
Rb$_3$(Cs$_3$)Sb$_2$(Bi$_2$)Br$_9$(I$_9$) perovskite compounds,
Inorganic Mat. (2011) 47, 2, 208-212.

\bibitem{Timmermans2011}
C. Timmermans, S. Cholakh, G. Blasse, The luminescence of
Cs$_3$Bi$_2$Cl$_9$ and Cs$_3$Sb$_2$Cl$_9$, J. Sol. State Chem. 46
(1983) 222-233.

\bibitem{Chabot1978}
B. Chabot, E. Parthe, Cs3Sb2I9 and Cs3Bi2I9 with the hexagonal
Cs3Cr2Cl9 structure type, Acta Cryst. B34 (1978) 645-648.

\bibitem{Sidey2000}
V. Sidey, Y. Voroshilov, S. Kun, E. Peresh, Crystal growth and
X-ray structure determination of Rb3Bi2I9, J. Alloys and Compounds
296 (2000) 53-58.

\bibitem{Lehner2013}
A. Lehner, D. Fabini, H. Evans, C. H?bert, S. Smock, J. Hu, H.
Wang, J. Zwanziger, M. Chabinyc, R. Seshadri, Crystal and
electronic structures of complex bismuth iodides A$_3$Bi$_2$I$_9$
(A = K, Rb, Cs) related to perovskite: Aiding the rational design
of photovoltaics, Chem. Mater. 27, 20 (2015) 7137-7148.

\bibitem{Preitschaft2004}
C. Preitschaft, ``Tern\"are und quatern\"are Materialien mit
komplexen Thio-, Selenido- und Halogenid- Anionen'', Dissertation,
Regensburg (2004).



\end{thebibliography}
\end{document}